\documentclass[useAMS,usenatbib]{mn2e}
\usepackage{graphics,graphicx}
\usepackage{aas_macros}
\usepackage{amssymb}
\usepackage{amsmath} 
\usepackage{latexsym}
\usepackage{color}
\usepackage{epstopdf}
\usepackage{marvosym}

\begin{document}

\title[The ortho-to-para ratio of water in interstellar clouds]{The
  ortho-to-para ratio of water in interstellar clouds}

%\title[The gas-phase nuclear-spin chemistry of interstellar water]{The
%  gas-phase nuclear-spin chemistry of interstellar water}
\author[A. Faure, P.  Hily-Blant, C. Rist, G. Pineau des For\^ets,
  A. Matthews and D.~R. Flower]{A. Faure$^{1}$\thanks{E-mail:
    alexandre.faure@univ-grenoble-alpes.fr}, P.
  Hily-Blant$^{1}$\thanks{E-mail:
    pierre.hily-blant@univ-grenoble-alpes.fr}, C.
  Rist$^{1}$\thanks{E-mail: claire.rist@univ-grenoble-alpes.fr},
  G. Pineau des For\^ets$^{2, 3}$\thanks{E-mail:
    guillaume.desforets@obspm.fr}, A. Matthews$^{1}$\thanks{E-mail:
    AliceeMaffs@live.co.uk} \newauthor and D.~R. Flower$^4$\thanks{E-mail:
    david.flower@durham.ac.uk} \\ $^1$ Univ. Grenoble Alpes, CNRS,
  IPAG, F-38000 Grenoble, France \\ $^2$ LERMA, UMR 8112 du CNRS,
  Observatoire de Paris, Ecole Normale Sup\'erieure, 61 Av. de
  l'Observatoire, 75014 Paris, France \\ $^3$  IAS, UMR8617 du CNRS, Universit\'e de Paris Sud 91705 Orsay France \\ $^4$ Physics Department, The
  University, Durham DH1 3LE, UK}

\date{Accepted ? Received ?}

\pagerange{\pageref{firstpage}--\pageref{lastpage}} \pubyear{2016}

\maketitle

\label{firstpage}

\begin{abstract}

The nuclear-spin chemistry of interstellar water is investigated using
the {\it University of Grenoble Alpes Astrochemical Network}
(\texttt{UGAN}). This network includes reactions involving the
different nuclear-spin states of the hydrides of carbon, nitrogen,
oxygen and sulphur, as well as their deuterated forms. Nuclear-spin
selection rules are implemented within the scrambling hypothesis for
reactions involving up to seven protons. The abundances and
ortho-to-para ratios (OPRs) of gas-phase water and water ions
(H$_2$O$^+$ and H$_3$O$^+$) are computed under the steady-state
conditions representative of a dark molecular cloud and during the
early phase of gravitational collapse of a prestellar core. The model
incorporates the freezing of the molecules on to grains, simple grain
surface chemistry and cosmic-ray induced and direct desorption of
ices. The predicted OPRs are found to deviate significantly from both
thermal and statistical values and to be independent of temperature
below $\sim $30~K. The OPR of H$_2$O is shown to lie between 1.5 and
2.6, depending on the spin-state of H$_2$, in good agreement with
values derived in translucent clouds with relatively high
extinction. In the prestellar core collapse calculations, the OPR of
H$_2$O is shown to reach the statistical value of 3 in regions with
severe depletion ($n_{\rm H}> 10^7$~cm$^{-3}$). We conclude that a low
water OPR ($\lesssim 2.5$) is consistent with gas-phase ion-neutral
chemistry and reflects a gas with OPR(H$_2)\lesssim 1$. Available OPR
measurements in protoplanetary disks and comets are finally discussed.

\end{abstract}


\section{Introduction}

The origin and distribution of water in the Solar System is not well
understood. In particular, the fraction of pristine i.e. unprocessed
interstellar water in comets and asteroids remains poorly
constrained. The contribution of comets and asteroids to the water
accreted by Earth is in turn a long-standing problem
\citep{alexander18}. Yet these questions are critical to understanding
star and planet formation in general and to assessing how typical is
the Solar System. Water is the second most abundant molecule (after
H$_2$) in the Solar System and is also an essential ingredient for
life on Earth.

Isotopic fractionation, i.e. the enrichment or depletion of an isotope
in a molecule relative to its elemental abundance, provides a powerful
diagnostic tool for tracing the chemical history of the Solar
System. For instance, the water D/H and $^{18}$O/$^{16}$O ratios in
the coma of comet 67P/Churyumov-Gerasimenko were measured recently by
the ROSINA mass spectrometer on board the {\it Rosetta}
spacecraft. Both ratios were found to be enriched with respect to the
terrestrial values, in agreement with the scenario where 67P's water
was inherited unprocessed from the presolar cloud
\citep{altwegg17,schroeder18}. In particular, the HDO/H$_2$O and
D$_2$O/HDO ratios, respectively $1.05\times 10^{-3}$ (more than three
times the terrestrial value) and $1.8\times 10^{-2}$, were found to be
similar to values reported for low-mass protostars embedded in
molecular clouds \citep[see][and reference
  therein]{altwegg17}. Additionally, disk models have shown that
unlike molecular clouds, the solar nebula protoplanetary disk was
probably unable to efficiently produce deuterium-rich water
\citep{cleeves14}. These studies thus suggest that a significant
fraction of the solar system's water is interstellar in origin (see
also \cite{vandishoeck14}).

%These studies are important for our understanding of the origin(s) of
%terrestrial oceans.

Likewise, the ortho-to-para ratio (OPR) of H$_2$O might be used as an
alternative tool to trace the link between interstellar, cometary and
planetary water. With its two equivalent hydrogen atoms, H$_2$O exists
in the form of two distinct nuclear-spin isomers, para ($I=0$, where
$I$ is the total nuclear-spin) and ortho ($I=1$), whose
interconversion in the gas-phase via radiative and inelastic
collisional transitions is forbidden or very slow. The statistical or
high-temperature OPR of H$_2$O is 3 and any OPR lower than 3 can be
interpreted in terms of an equilibrium {\it spin temperature}. In
comets, the spin temperature has been traditionally considered as a
proxy for the formation temperature of water ice
\citep{mumma86,bonev13}. Typical values for the OPR of H$_2$O in
comets lie in the range $2-3$, corresponding to spin temperatures
lower than 50~K \citep[see][and references therein]{faggi18}. In the
interstellar medium (ISM), it is generally believed that the OPR of
water formed in the gas-phase should be statistical i.e. equal to
3. The OPR of H$_2$O in the ISM was accurately measured using the HIFI
spectrometer on board the {\it Herschel} space observatory. In diffuse
and translucent clouds the OPR of H$_2$O is usually consistent with
the statistical value, but values in the range $2-3$ have also been
reported \citep[see][and references therein]{vandishoeck13}. In
protoplanetary disks, an accurate measure is not available but
estimates are consistent with the interstellar and cometary range of
$2-3$ \citep{pontoppidan10,salinas16}. In the cold and dense ISM, only
the ground-state oH$_2$O transition was detected towards the L1544
prestellar core \citep{caselli12}, thus precluding a measure of the
OPR.

In summary, many H$_2$O OPR measurements are now available for comets
and interstellar clouds, but their meaning remains unclear. In
addition, the above assumptions linking the spin-state of a molecule
to its formation process have been recently challenged by both
experiment and theory. First, the OPR of H$_2$O photodesorbed and
thermally desorbed from ice at 10~K and 150~K, respectively, was found
equal to the statistical value of 3, even when the ice was produced
{\it in situ} at 10~K \citep{hama16} and when the ice was made from
pH$_2$O monomers \citep{hama18}. The assumed relation between the OPR
and the formation temperature of water ice is thus not supported
experimentally. Second, the existence of nuclear-spin selection rules
in chemical reactions, as predicted theoretically by \cite{quack77},
has been demonstrated experimentally in ion-neutral reactions
involving H$_3^+$ \citep{uy97,crabtree11b}. The nuclear-spin chemistry
of interstellar molecules has gained interest in recent years and
detailed models have been dedicated to the OPR of NH$_3$
\citep{faure13,legal14}, H$_3^+$ and its deuterated isotopologues
\citep{albertsson14,harju17b}, deuterated ammonia \citep{harju17},
H$_2$O$^+$ \citep{herbst15} and H$_2$Cl$^+$
\citep{neufeld15,legal17}. These studies have shown that the OPR of
molecules formed in the gas-phase can be significantly lower than the
statistical (high-temperature limit) values and is entirely controlled
by chemical selection rules.

%These selection rules were found to explain, in particular, the
%anomalous OPR of NH$_3$ ($\sim 0.5$) measured with {\it Herschel} in
%cold ($T\sim 30$~K) interstellar clouds \citep{persson12}. The main
%result is that the OPR of molecules formed {\it via} gas-phase H$_2$
%abstractions are not predicted to be statistical but to be entirely
%driven by the OPR of H$_2$ \citep{faure13}. Other gas-phase processes
%such as proton-exchange reactions between H$_2$O and abundant ions
%(e.g. H$^+$ and H$_3$O$^+$) have been also invoked to explain the
%``anomalous'' OPRs of H$_2$O below 3 \citep{vandishoeck13,hama18}.

In this work, we investigate the nuclear-spin chemistry of gas-phase
water from interstellar clouds ($T\leq 100$~K) to cold prestellar
cores ($T\sim 10$~K). Our model is based on the {\it University of
  Grenoble Alpes Astrochemical Network} (\texttt{UGAN}) as recently
published by \cite{hily18} (hereafter HB18). This network includes the
nuclear-spin states of H$_2$, H$_2^+$, H$_3^+$ and of all the hydrides
of carbon, nitrogen, oxygen and sulphur, as well as their abundant
deuterated forms. It was used by HB18 to study the deuterated
isotopologues of ammonia in collapsing pre-stellar sources.

In Section~2, we summarize the update of the \texttt{UGAN} network for
the water chemistry. Section~3 contains our results both for the
steady-state composition of a molecular cloud with uniform density and
temperature and for a collapsing core. Comparison with available
observations is discussed in Section~4 and Section~5 gives our
concluding remarks.

\section{The models}

The aim of the present work is to compute the abundance and OPR of
H$_2$O (and its precursors) under the steady-state conditions
representative of a dark molecular cloud and also during the inital
stage of gravitational collapse of a prestellar core, which ultimately
leads to the formation of a low-mass protostar. The dynamical model
was presented in HB18 and derives from the studies of gravitational
collapse by \cite{larson69} and \cite{penston69}. Briefly, the
collapsing core loses mass to the surrounding envelope at a rate that
ensures that the density profile in the envelope is proportional to
$R_{env}^{-2}$, where $R_{env}$ is the envelope radius. Full details
can be found in HB18. This dynamical model is combined with the
\texttt{UGAN} chemical network, also presented in HB18, which is an
upgraded version of the gas-phase network of \cite{flower06}
(hereafter F06). The F06 network included reactions involving species
containing H, D, He, C, N, O and S and distinguished between the
different nuclear-spin states of H$_2$, H$_2^+$, H$_3^+$ (including
deuterated forms) and between those of nitrogen hydrides. A first
update of the F06 network consisted in a revision of the
nitrogen-hydrides' chemistry (excluding deuterated species)
\citep{legal14}. In particular, the nuclear-spin selection rules were
derived with the method of \cite{oka04}, which is based on the
conservation of the rotational symmetry of the nuclear-spin
isomers. These symmetry rules depend upon the mechanism of reaction,
and two extreme mechanisms can be considered: hopping and
scrambling. In HB18, scrambling was assumed because at very low
temperature, ion-neutral reactions usually form long-lived
intermediate complexes in which complete randomization of H and/or D
atoms can take place, as shown experimentally \citep{crabtree11b}. The
second update by HB18 consisted of extending the work of
\cite{legal14} to the entire F06 network in a systematic fashion for
all hydrides containing C, N, O and S atoms, and their deuterated
forms. To this end, the nuclear-spin selection rules were derived from
the permutation symmetry approach of \cite{quack77}, which is more
general and adapted to deuterium nuclei. The nuclear-spin separation
procedure is described in detail in HB18. Finally, many reaction rate
coefficients were updated from a literature survey.

We describe below the third update of the F06 network which mainly
consists of a revision of the oxygen hydrides chemistry.

\subsection{Water chemistry}

The chemistry of interstellar water can follow three distinct routes
\citep{vandishoeck13}: {\it i)} low-temperature ion-neutral gas-phase
chemistry ($T\leq 100$~K), {\it ii)} high-temperature neutral-neutral
gas-phase chemistry and {\it iii)} surface chemistry. In the present
work, the first and third routes are included but surface reactions
are treated in a very simple fashion (see Section~2.2 below).

The low temperature ion-neutral synthesis of H$_2$O starts with the
ionization of H$_2$ by cosmic-ray protons and secondary
electrons. This leads to H, H$^+$, and H$_2^+$ and also to H$_3^+$ via
the fast reaction between H$_2$ and H$_2^+$. Oxygen atoms react with
either H$^+$ to create O$^+$ ions (by charge-transfer) or with H$_3^+$
to form OH$^+$ and H$_2$O$^+$:
\begin{equation}
  {\rm O + H^+ \to O^+ + H}
  \label{eq1}
\end{equation}
\begin{equation}
  {\rm O^+ + H_2 \to OH^+ + H}
  \label{eq2}
\end{equation}
\begin{equation}
  {\rm O + H_3^+ \to OH^+ + H_2 \;or\; H_2O^+ + H}
  \label{eq3}
\end{equation}
Water is then formed via a small chain of exothermic and barrierless
reactions:
\begin{equation}
  {\rm OH^+ + H_2 \to H_2O^+ + H}
  \label{eq4}
\end{equation}
\begin{equation}
  {\rm H_2O^+ + H_2 \to H_3O^+ + H}
    \label{eq5}
\end{equation}
\begin{equation}
  {\rm H_3O^+ + e^- \to H_2O + H}
    \label{eq6}
\end{equation}
It should be noted that because the formation of H$^+$ and H$_3^+$ is
initiated by the cosmic-ray ionization of H$_2$, the formation of
OH$^+$, H$_2$O$^+$ and H$_3$O$^+$ is essentially cosmic-ray driven and
their relative abundance can be used to constrain the cosmic-ray
ionization rate \citep{hollenbach12,indriolo15}.

The rate coefficients for reactions~(\ref{eq1}), (\ref{eq3}) and
(\ref{eq6}) are those of HB18. The rate coefficient for
reaction~(\ref{eq1}), averaged over the fine-structure levels of
oxygen, is taken from the drift-tube measurements of \cite{federer84}
at 300~K. Theoretical calculations by \cite{chambaud80},
\cite{stancil99} and \cite{spirko03} are in good agreement and they
all lie within the experimental error bar ($\pm 50$\%). However, the
theoretical results differ significantly from each other at the
state-to-state level, especially when the oxygen atom is in the ground
state ($^3P_2$). State-resolved experimental data below 100~K would be
very useful to resolve the disagreement between the calculations. This
is particularly important where most oxygen atoms are in the ground
state The rate coefficient for reaction~(\ref{eq3}) is taken from the
transition state theory calculations of \cite{klippenstein10} and it
was combined with the experimental branching ratios determined by
\cite{milligan00} at 300~K. The rate coefficient and branching ratios
for reaction~(\ref{eq6}) are taken from the storage ring measurements
of \cite{jensen00}, which agree well with the most recent results of
\cite{buhr10} for D$_3$O$^+$.

The rate coefficients for reactions~(\ref{eq2}), (\ref{eq4}) and
(\ref{eq5}) have been updated using very recent experimental studies
performed at low temperatures. Reaction~(\ref{eq2}) was studied by
\cite{kovalenko18} in a ion trap down to 15~K and it was found to be
almost temperature independent. The rate coefficient is taken here as
the value measured at 15~K, i.e. $1.3\times 10^{-9}$~cm$^3$s$^{-1}$,
with no temperature dependence.
%We note that the wavepacket calculations of \cite{bulut15} predict a
%slightly steeper temperature dependence.
Reactions~(\ref{eq4}) and (\ref{eq5}) were studied in ion traps by
both \cite{tran18} and \cite{kumar18}. Again the temperature
dependence was found to be small and the rate coefficients are taken
as the values measured at 21~K by \cite{kumar18} i.e. $1.22\times
10^{-9}$~cm$^3$s$^{-1}$ for reaction~(\ref{eq4}) and $1.57\times
10^{-9}$~cm$^3$s$^{-1}$ for reaction~(\ref{eq5}), with no temperature
dependence. These values are in good agreement with ring polymer
calculations performed by \cite{kumar18} and agree also with the
independent measurements of \cite{tran18}, within typically
30\%. These new measurements confirm that O$^+$, OH$^+$ and H$_2$O$^+$
ions react very fast with molecular hydrogen down to interstellar
temperatures, with rate coefficients close to the Langevin limit.

Finally, the above rate coefficients were duplicated to consistently
update the deuterated homologues of reactions~(1-6). In this
``cloning'' procedure, the overall rate coefficients of the deuterated
reactions are assumed to be the same as the original (hydrogenated)
reactions, except when isotope measurements are available. This point
is further discussed in Section~3.1.4 below.

\subsection{Grain surface processes}

Grain-surface reactions are not explicitly included in the
\texttt{UGAN} network, except the formation (and immediate desorption)
of H$_2$ and isotopologues. The rates of adsorption of neutral species
include the contribution of the ice mantle thickness to the grain
cross section, as described in the Appendix~B of
\cite{walmsley04}. All oxygen atoms from the neutral species O, OH,
H$_2$O, NO, SO and SO$_2$ are assumed to form water ice once they are
adsorbed by the grains. The list of species formed in grain mantles is
given in the Table~D1 of HB18.

The desorption of molecules by the cosmic-ray induced ultraviolet
radiation field is included and described in the Appendix~A of
HB18. Our treatment of desorption induced by direct cosmic-ray impact
follows the formulation of \cite{flower05} (see their
Section~3.3). Briefly, the rate of desorption of species $i$ (averaged
over the cosmic-ray flux) per unit volume per unit time from the
grains is:
\begin{equation}
  R_i^{crd}=\frac{n_i^g}{\sum_in_i^g}n_g\pi a_g^2\gamma_{\rm
    CO}\exp\left[\frac{-(E_i^{\rm ads}-E_{\rm CO}^{\rm ads})}{T_g^{\rm
        max}}\right]
\label{crd}
\end{equation}
where $\frac{n_i^g}{\sum_in_i^g}$ is the fractional abundance of
species $i$ on grains, $n_g$ is the number density of grains of radius
$a_g$, $\gamma_{\rm CO}$ is the CO yield averaged over the cosmic-ray
flux (in molecules~cm$^{-2}$~s$^{-1}$), $E_i^{\rm ads}$ is the
adsorption energy of the species (as a pure ice) and $T_g^{\rm max}$
is the maximum temperature reached by the grains following cosmic-ray
impact. This formulation is similar (but simpler) than that of
\cite{hasegawa93}. It assumes, in particular, an exponential
dependence of the desorption rate on adsorption energy, as for thermal
evaporation. Following Flower et al. (2005), we adopted $T_g^{\rm
  max}$=70~K, as derived by \cite{hasegawa93}. For the CO yield (in
~molecules~cm$^{-2}$s$^{-1}$) we used $\gamma_{\rm CO}=70~\zeta_{17}$
where $\zeta_{17}$ is the rate of cosmic-ray ionization of molecular
hydrogen in unit of $10^{-17}$s$^{-1}$. This value for $\gamma_{\rm
  CO}$ was derived by \cite{leger85} for ``spot'' heating
(i.e. sputtering) of grain mantles. In the case of species with high
adsorption energies like water and ammonia, however, the above
formulation underestimates the desorption rates by orders of magnitude
\citep{bringa04}. In fact, the desorption rate for such species is
dominated by the ``prompt'' or very early desorption which does not
scale as $\exp(-E_i^{\rm ads})$. \cite{bringa04} have suggested a
scaling $R_i^{crd}\propto (E_i^{\rm ads})^{-m}$ with $m\sim 2$. For
H$_2$O (and isotopologues), $\gamma_{\rm H_2O}$ was thus directly
computed from experimental data (see Appendix~A). We obtained
$\gamma_{\rm H_2O}= 0.8~\zeta_{17}$, i.e. about a factor of 100
smaller than $\gamma_{\rm CO}$. For NH$_3$ (and isotopologues), we
adopted the same formulation as for H$_2$O with $\gamma_{\rm
  NH_3}=(E_{\rm H_2O}^{\rm ads}/E_{\rm NH_3}^{\rm ads})^2\gamma_{\rm
  H_2O}=2.96~\gamma_{\rm H_2O}$~molecules~cm$^{-2}$s$^{-1}$ using
binding energies from \cite{brown07}. For all other species,
Eq.~(\ref{crd}) was employed.

%We note that new measurements for the sputtering of CO are
%in progress (Dartois, private communication).

%In addition, the above formalism does not take into account the ice
%mantle thickness dependence. This parameter was experimentally shown
%to play a role for grain mantles with less than $\sim 30$ layers
%\citep{dartois18}, i.e. in interstellar regions with visual
%extinctions $A_v \lesssim 3$.

%It should be noted that the above formulations and yields are
%expected to provide the right orders of magnitude but large
%uncertainties in the experimental data and in the cosmic-ray spectra
%preclude precise estimates (see Appendix A).

Taking typical conditions for dark molecular clouds ($n_{\rm
  H}=10^4$~cm$^{-3}$, $\zeta=3\times 10^{-17}$~s$^{-1}$,
$n_g=1.7\times 10^{-8}$~cm$^{-3}$, $\sum_in_i^g=1.9$~cm$^{-3}~$,
$a_g=0.13~\mu$m), the desorption rates per unit time for CO and H$_2$O
are $k_{\rm CO}^{crd}\sim 9.5\times 10^{-16}$~s$^{-1}$ and $k_{\rm
  H_2O}^{crd}\sim 1.1\times 10^{-17}$~s$^{-1}$, which agree within a
factor of $\sim 2-4$ with the ``experimental'' rates derived by
\cite{bringa04}. The corresponding time scales are $t_{\rm CO}^{crd}
\sim 3.3\times 10^7$~yr and $t_{\rm H_2O}^{crd} \sim 2.3\times
10^9$~yr, which are both longer than the typical lifetime of a
molecular cloud. We note that $t_{\rm H_2O}^{crd}$ is very similar to
the time scale for cosmic-ray induced photodesorption ($\sim 10^9$~yr,
see \cite{hily18}), meaning that the two processes are in competition
and will become significant at the higher densities of prestellar
cores (see Section~3.2).

Finally, a gas-phase OPR equal to the statistical (high-temperature)
value is assumed for all species upon cosmic-ray (induced and direct)
desorption, as suggested by the photodesorption experiments of Hama et
al. (2018). Thus, whatever the mechanism, a water molecule desorbed
from ice will have an initial OPR of 3 once in the gas-phase.

\section{Results}

The initial distribution of the elements is specified in Tables~2 and
3 of HB18. In particular, the fractional abundance of oxygen atoms
(relative to the total H nuclei density $n_{\rm H}$) in the gas-phase
is $1.24\times 10^{-4}$ and that of H$_2$O molecules in the grain
mantles is $1.03\times 10^{-4}$. A discussion on the uncertainties
surrounding the elemental abundances can be found in \cite{legal14}.

\subsection{Steady-state composition}

We first investigate the steady-state abundances of oxygenated species
for an interstellar cloud having a uniform density ($n_{\rm
  H}=10^4$~cm$^{-3}$) and kinetic temperature ($T$) in the range
$5-100$~K. The cosmic-ray ionization rate of H$_2$ is taken as
$\zeta=3\times 10^{-17}$~s$^{-1}$ in our reference model, which is
close to the average rate inferred from molecular ion observations in
dense clouds \citep[see][and references therein]{indriolo12}. We have
also studied the impact of a larger rate, $\zeta=3\times
10^{-16}$~s$^{-1}$, more representative of diffuse or translucent
clouds \citep{neufeld17}. The initial radius of the refactory grain
core is taken as 0.1~$\mu$m. It should be noted that in these
steady-state runs, grain surface processes are turned off so that the
mantle composition (see Table~3 of HB18) is fixed and most of the
oxygen is locked into ices (mainly water). We have also ignored the
presence of an external far-ultraviolet (FUV) field so that the
results below are most relevant for interstellar clouds with moderate
to high extinction ($\gtrsim$3~mag.), i.e. from translucent to dark
cloud conditions. In particular, the molecular hydrogen fraction
($f_{\rm H_2}=2n({\rm H_2})/n_{\rm H}$) is close to unity and the
electron fraction is lower than $3\times 10^{-7}$ in our
simulations. The temperature is varied from 5 to 100~K to explore a
large range of OPRs of H$_2$ (see below).

Table~1 presents the steady-state abundances of several species
related to H$_2$O at 10~K and for two values of $\zeta$. We first
notice that the fractional abundance of H$_2$O is $\sim 3.5\times
10^{-7}$. This value is similar to the H$_2$O peak abundance in the
elaborate photodissociation region (PDR) models of
\cite{hollenbach09}. In these models, the water peak occurs at visual
extinctions $A_V\sim 3-8$, depending on the incident FUV field. When
averaged through the cloud, the H$_2$O abundance becomes $\sim
10^{-8}$ \citep{hollenbach09}, in good agreement with observations of
diffuse and translucent clouds \citep[see][and references
  therein]{vandishoeck13}. In our model, when $\zeta=3\times
10^{-16}$~s$^{-1}$, the abundances of OH$^+$, H$_2$O$^+$ and
H$_3$O$^+$ are lower but within a factor of $3-10$ of those in
\cite{hollenbach12} at the `second' peak, i.e. $A_V\sim 5$ (see their
Fig.~5).

In our model the abundances of O, O$_2$, H$_2$O and H$_3$O$^+$ do not
strongly vary with $\zeta$ (see Table~1). In contrast, the abundances
of OH$^+$ and H$_2$O$^+$ are found to scale roughly linearly with
$\zeta$ (similarly to H$_3^+$). This was previously discussed by
\cite{hollenbach12}. As a direct consequence, the OH abundance is
multiplied by a factor of $\sim 5$, decreasing the H$_2$O/OH ratio
from 2 to 0.4. In diffuse and translucent clouds, this ratio lies in
the range $0.3-1$ \citep{wiesemeyer12}. We note that our model
predicts that most of the gas-phase oxygen (apart from CO) is in O and
O$_2$. The predicted abundance of O$_2$ is significantly larger than
the observed values, which is an old problem in astrochemistry
\citep{goldsmith11}.

\begin{table}
\caption{Steady-state abundances expressed relative to $n_{\rm
    H}=n({\rm H})+2n({\rm H_2})=10^4$~cm$^{-3}$. The kinetic
  temperature is fixed at 10~K.  Two values of the cosmic-ray
  ionization rate $\zeta$ (in s$^{-1}$) have been used. Numbers of
  parentheses are powers of 10.}
\begin{center}
\begin{tabular}{lcc}
\hline Species & $\zeta=3\times 10^{-17}$ & $\zeta=3\times 10^{-16}$\\ \hline
H           & 6.9(-05)  &  7.6(-04)     \\
pH$_2$      & 5.0(-01)  &  5.0(-01)     \\
oH$_2$      & 5.4(-04)  &  8.7(-04)     \\
pH$_3^+$    & 4.1(-09)  &  3.0(-08)     \\
oH$_3^+$    & 1.8(-09)  &  1.5(-08)     \\
O           & 2.1(-05)  &  2.9(-05)     \\
OH          & 1.7(-07)  &  8.6(-07)     \\
O$_2$       & 9.6(-06)  &  6.1(-06)     \\
pH$_2$O     & 1.4(-07) &  1.5(-07)     \\
oH$_2$O     & 2.1(-07)  &  2.2(-07)     \\
OH$^+$      & 3.2(-13)  &  3.2(-12)     \\
pH$_2$O$^+$ & 1.6(-13)  &  1.6(-12)     \\
oH$_2$O$^+$ & 1.8(-13)  &  1.9(-12)     \\
pH$_3$O$^+$ & 2.3(-09)  &  4.3(-09)     \\
oH$_3$O$^+$ & 5.6(-10)  &  1.0(-09)     \\
e$^-$       & 4.2(-08)  &  2.2(-07)     \\ \hline
\end{tabular}
\end{center}
\label{table:1}
\end{table}

The OPRs of the nuclear-spin isomers listed in Table~1 are given in
Table~2. We can first observe significant deviations from thermal
values. The OPR of H$_2$, in particular, is suprathermal and
corresponds to a spin temperature of $\sim 20$~K. We note that the
value predicted for H$_2$O$^+$ (1.2) is forbidden in thermal
equilibrium because the thermal OPR of H$_2$O$^+$ is necessarily
larger than 3. This is analogous to the case of NH$_3$ whose OPR is
predicted to be lower than unity at low temperature
\citep{faure13}. The OPRs of water and its ions are found to be
insensitive to $\zeta$. We have checked that they are also insensitive
to the gas-phase abundance of sulphur, which controls the fractional
ionization when $\zeta$ is fixed.

\begin{table}
\caption{Steady-state OPRs of the nuclear-spin isomers listed in
  Table~1. The kinetic temperature is fixed at 10~K.  Two values of
  the cosmic-ray ionization rate $\zeta$ (in s$^{-1}$) have been
  used. Numbers of parentheses are powers of 10. Thermal (at 10~K) and
  statistical OPRs are also provided.}
\begin{center}
\begin{tabular}{lcccc}
  \hline Species & $\zeta=3\times 10^{-17}$ & $\zeta=3\times 10^{-16}$ & Thermal & Stat. \\ \hline  
H$_2$   & 1.1(-03)  &  1.8(-03) & 3.5(-07) & 3    \\
H$_3^+$ & 0.43      &  0.52     & 0.075    & 1  \\
H$_2$O  & 1.5       &  1.5      & 0.31    & 3 \\
H$_2$O$^+$  & 1.2   &  1.2      & 19      & 3 \\
H$_3$O$^+$  & 0.24   &  0.24    & 0.94    & 1 \\ \hline
\end{tabular}
\end{center}
\label{table:2}
\end{table}

\subsubsection{Temperature dependence}

In Fig.~1, the OPRs of H$_2$ and H$_3^+$ are plotted as function of
the kinetic temperature. Below $\sim 20$~K, these ratios are
suprathermal and almost independent of temperature: the OPR of H$_2$
is $\sim 10^{-3}$ and that of H$_3^+$ $\sim 0.4$. In this temperature
regime, the formation of oH$_2$ on the grains is faster than the
gas-phase conversion from pH$_2$ to oH$_2$ (due to proton exchanges
with H$^+$, H$_3^+$ and HCO$^+$). A ``critical'' temperature $T_{\rm
  crit}$ was defined in \cite{faure13} to quantitatively explain this
effect (see their Eq.~(6)). Thus, above $T_{\rm crit}\sim 20$~K, the
OPR of H$_2$ rapidly reaches thermal equilibrium because the
nuclear-spin conversion rate of H$_2$ becomes faster than the
formation rate. It should be noted that for the equilibration reaction
H$_3^+$+H$_2$, we have adopted {\it species-to-species} rate
coefficients, as computed by HB18 from the data of \cite{hugo09} with
the assumption that rotational populations are at local thermal
equilibrium (LTE). As a result, above 20~K, not only the OPR of H$_2$
but also that of H$_3^+$ reaches thermal equilibrium in our model. In
fact, small deviation can be seen at high-temperatures because the
state-to-state data of Hugo et al. (2009) are strictly valid up to
50~K only.

Is is instructive to compare these results with values inferred from
infrared and ultraviolet absorption observations. The OPR of H$_2$ has
been measured in diffuse clouds with values ranging from $\sim$ 0.3 to
1.5, i.e. spin temperatures between 50 and 100~K \citep[see][and
  references therein]{crabtree11}. In dense clouds, H$_2$ is very
difficult to detect and, to our knowledge, the only (published) direct
measurements are upper limits reported by \cite{lacy94} towards
NGC~2024 (OPR$<0.8$) and by \cite{goto15} towards NGC~7538~IRS~1
(OPR$<2.3$), which are consistent with indirect estimates \citep[see
  e.g.][and references therein]{troscompt09,dislaire12}. The OPR of
H$_3^+$ has been measured in both dense \citep{mccall99} and diffuse
clouds \citep{crabtree11} with values in the range $0.4-1$,
corresponding to spin temperatures $\sim 20-50$~K for both types of
clouds. In diffuse clouds where both H$_2$ and H$_3^+$ have been
detected, the average H$_2$ spin temperature is $\sim$60~K while that
of H$_3^+$ is $\sim 30$~K \citep{crabtree11}. This discrepancy is
puzzling because both species are expected to be thermalized at these
temperatures, just as in our model (see Fig.~1). \cite{crabtree11}
have shown that in contrast to H$_2$, the OPR of H$_3^+$ is likely to
be non-thermal in diffuse clouds. However, the spin-state of H$_3^+$
plays only a minor role in the spin-chemistry of H$_2$O, as shown
below.

\begin{figure}
\includegraphics*[width=7.5cm,angle=-90.]{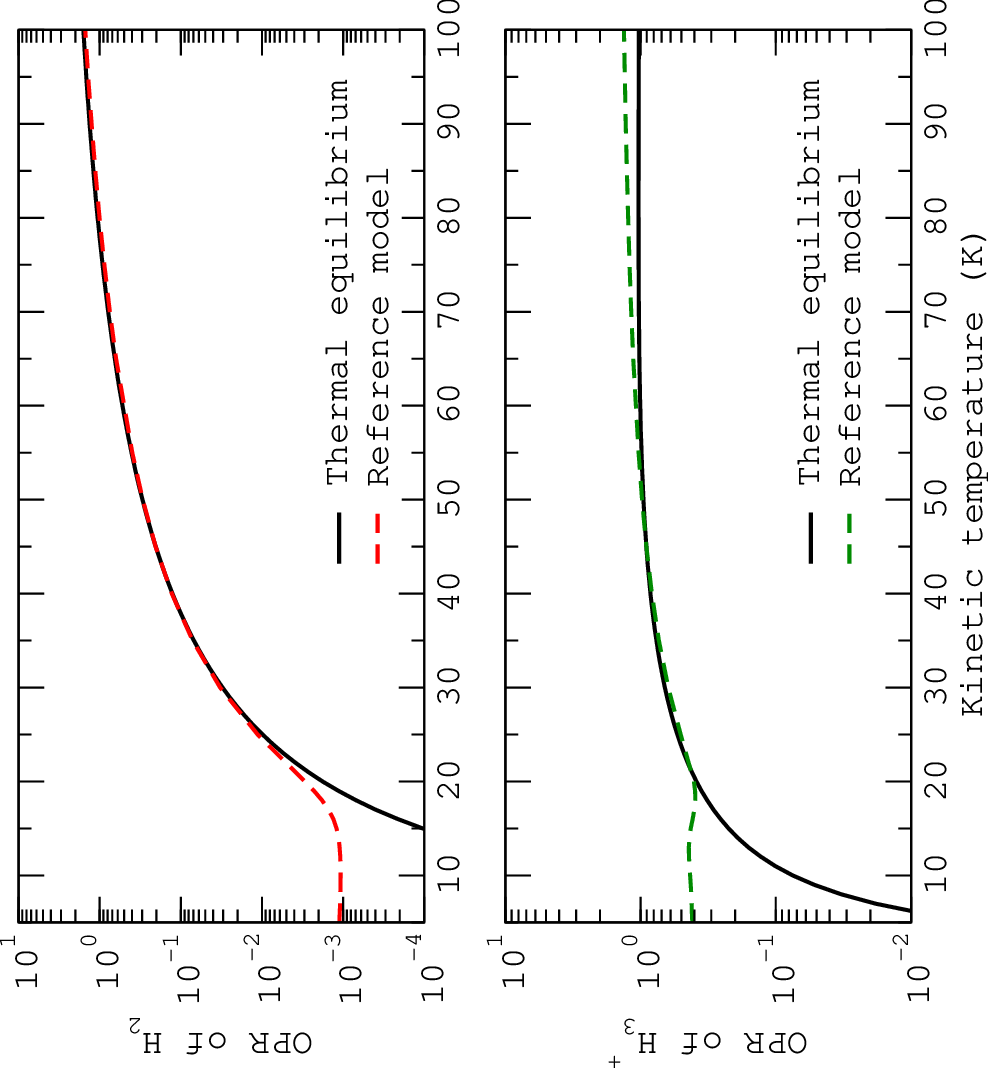}
\caption{OPRs of H$_2$ and H$_3^+$ as function of kinetic
  temperature. The solid lines give the thermalized OPRs. The dashed
  lines correspond to our reference model.}
\label{fig1}
\end{figure}

In Fig.~2, the same plot is given for water and the water ions
H$_2$O$^+$, H$_3$O$^+$. We find that the predicted OPRs deviate
significantly from thermal values over the whole temperature range,
i.e. $5-100$~K. We can also observe that these OPRs are independent of
temperature below $\sim$30~K. This result is reminiscent of the work
of \cite{faure13} for the nuclear-spin chemistry of ammonia. As
explained by these authors, the gas-phase OPR of NH$_3$ is driven by
the OPR of H$_2$ because its direct precursor, NH$_4^+$, is formed
through a series of hydrogen abstraction reactions with
H$_2$. Similarly here, the direct precursor of water, H$_3$O$^+$, is
mostly formed via reactions with H$_2$ (see reactions~(2), (4) and
(5)). Above 30~K, the OPRs of H$_2$O$^+$, H$_3$O$^+$ and H$_2$O
steadily increase towards their thermal value, although they do not
reach it in the explored temperature range. An important finding is
that the OPR of H$_2$O never goes below 1.5 and that it is subthermal
in the range $20-100$~K, with values comprised between 1.5 and
2.6. Comparisons with the observational OPRs of H$_2$O$^+$, H$_3$O$^+$
and H$_2$O will be presented in Section~4.

\begin{figure}
\includegraphics*[width=7.5cm,angle=-90.]{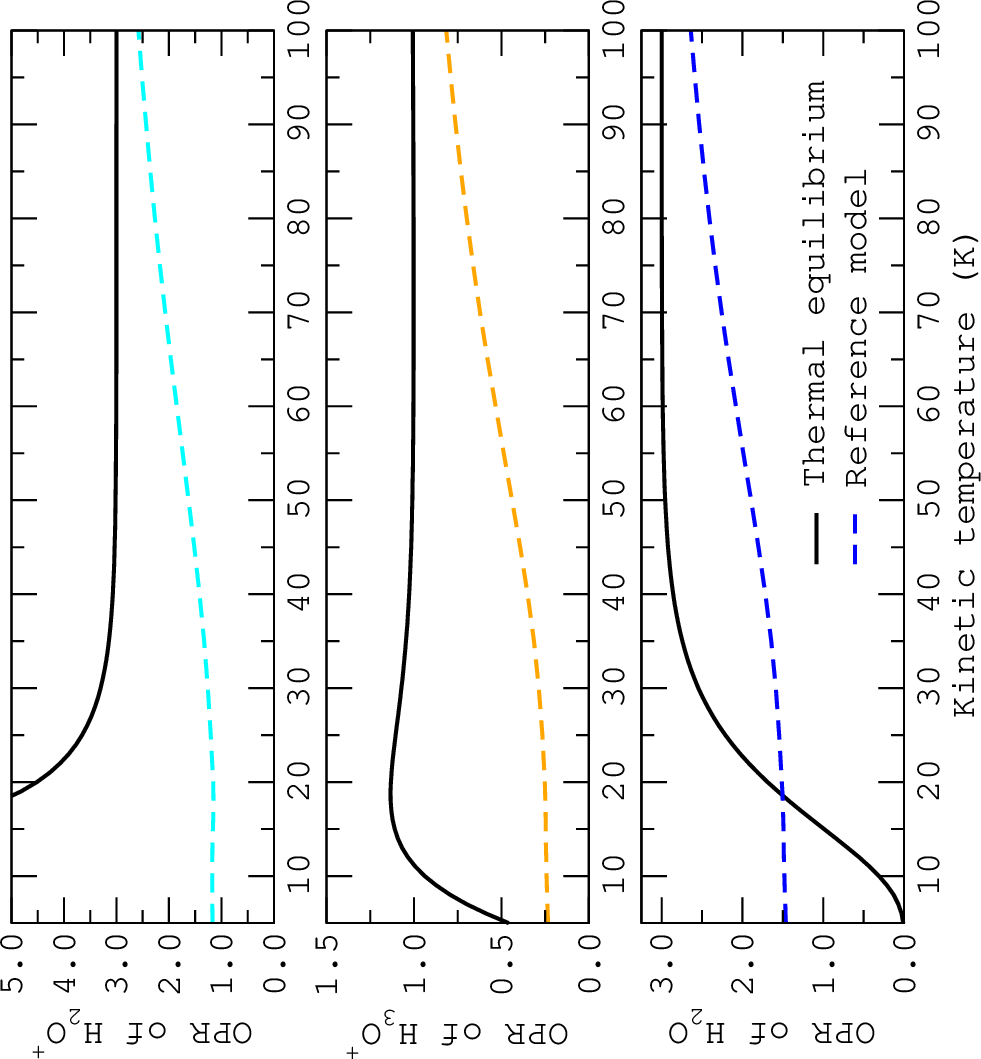}
\caption{OPRs of H$_2$O$^+$, H$_3$O$^+$ and H$_2$O as function of
  kinetic temperature. The solid lines give the thermalized OPRs. The
  dashed lines correspond to our reference model. }
\label{fig2}
\end{figure}

\subsubsection{H$_2$ OPR dependence}

It is now instructive to combine the results of Figs.~1 and 2 in order
to plot the OPRs of water and water ions as function of the OPR of
H$_2$. As shown in Fig.~3, the quantity of oH$_2$ has no impact as
long as the OPR of H$_2$ is lower than $\sim$ 0.1. Indeed, in this
regime, the formation of the water ions follows hydrogen abstractions
in a {\it para-rich} H$_2$ gas. The abundance of oH$_2$ starts to play
a role and the OPRs of water and its ions increase significantly only
when OPR(H$_2$) $\gtrsim 0.1$. This can be understood analytically by
deriving the OPRs from the nuclear-spin branching ratios of
reactions~(3-6), as explained in Appendix~B. Assuming that the
reaction of H$_3^+$ with oxygen atoms is a negligible source of
H$_2$O$^+$, we obtain the analytic results plotted in Fig.~3 as
dotted-lines. The good agreement between our reference model and the
analytic calculation demonstrates that the OPR of H$_2$O is governed
by H$_2$ abstractions and, consequently, by the OPR of H$_2$. The OPR
of H$_3^+$ is thus found to play a minor role in the nuclear-spin
chemistry of H$_2$O. The reaction of H$_3^+$ with O is, however, a
significant source of OH$^+$.

\begin{figure}
\includegraphics*[width=7.5cm,angle=-90.]{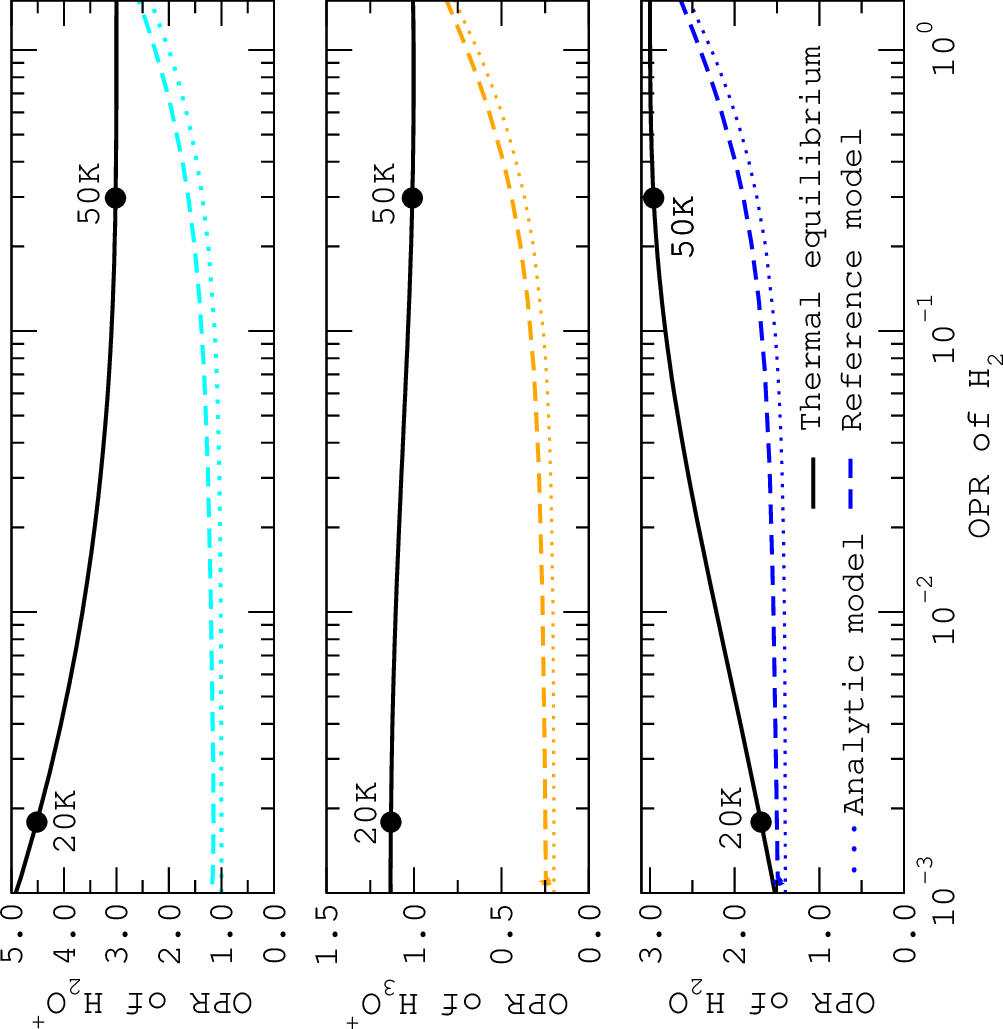}
\caption{OPRs of H$_2$O$^+$, H$_3$O$^+$ and H$_2$O as function of the
  OPR of H$_2$. The solid lines give the thermalized OPRs. The dashed
  lines correspond to our reference model. The analytical model
  described in Appendix~B is denoted by the dotted lines. Values of
  the kinetic temperature increase from left to right and are
  indicated at 20 and 50~K.}
\label{fig3}
\end{figure}

\subsubsection{Thermalization processes}

A number of thermoneutral equilibration (i.e. thermalization)
processes can influence the steady-state OPR of water and its
ions. These equilibration processes are neglected in the \texttt{UGAN}
network, except the reactions of H$_2$ with H$^+$, H$_3^+$ and HCO$^+$
(and their deuterated isotopologues). We investigate in this section
the impact of other possible equilibration reactions on the OPR of
H$_2$O$^+$, H$_3$O$^+$ and H$_2$O. The kinetic temperature is varied
in the range 5-50~K.

For H$_2$O$^+$, a possible equilibration reaction is:
\begin{equation}
  {\rm pH_2O^+ + H \leftrightarrows oH_2O^+ + H}.
\label{h2opeq}
\end{equation}
It was indeed considered by \cite{herbst15} in his study of the OPR of
H$_2$O$^+$ in diffuse clouds. In these environments,
reaction~(\ref{h2opeq}) can be faster than the (dominant) destruction
reaction with electrons because hydrogen atoms are relatively
abundant. In our reference model, however, the molecular hydrogen
fraction is near unity ([H]/[H$_2$]$\sim 10^{-4}$) and
reaction~(\ref{h2opeq}) is expected to be of minor importance. We have
added this reaction to our network assuming that it proceeds with a
rate coefficient of $2\times 10^{-9}$~cm$^3$s$^{-1}$ in the exothermic
direction, i.e. near the capture (Langevin) limit. The detailed
balance principle was applied for the reverse endothermic channel. The
OPR of H$_2$O$^+$ was found to change by less than 0.1\%, meaning that
reaction~(\ref{h2opeq}) is entirely negligible with our physical
conditions.

Once water is formed, a similar equilibration reaction is:
\begin{equation}
  {\rm pH_2O + H^+ \leftrightarrows oH_2O + H^+},
\label{h2oeq}
\end{equation}
as suggested by Hama et al. (2018). The reaction of H$_2$O with
protons, however, produces H$_2$O$^+$ via a strongly exothermic charge
transfer with a rate coefficient of $8.2\times 10^{-9}$cm$^3$s$^{-1}$
at room temperature \citep{huntress74}. The occurrence of proton
exchange in the intermediate complex H$_3$O$^+$ is therefore
uncertain. Indeed, to be efficient the scrambling of hydrogen requires
a sufficiently long lifetime of the intermediate complex so that many
vibrations can occur before dissociation. Owing to the fast charge
transfer process, scrambling is not expected to occur and
reaction~(\ref{h2oeq}) was not considered.

Water can still exchange protons via the reaction with H$_3$O$^+$:
\begin{equation}
  {\rm pH_2O + (o, p)H_3O^+ \leftrightarrows oH_2O + (o, p)H_3O^+}.
  \label{h3opeq}
\end{equation}
Deuterated variants of this reaction were indeed studied
experimentally by \cite{smith80} at 300~K and isotope H/D exchanges
were observed. The overall rate coefficient was measured as $\sim
2\times 10^{-9}$~cm$^3$s$^{-1}$, i.e. close to the capture limit in
the average-dipole-orientation (ADO) approximation. In addition, the
product distributions were found to be purely statistical, implying
that the reaction proceeds via the formation of an intermediate
long-lived complex. We have estimated the species-to-species rate
coefficients for reaction~(\ref{h3opeq}) by combining the capture ADO
value at 10~K ($\sim 10^{-8}$~cm$^3$s$^{-1}$) with the simple
statistical model of \cite{rist13}. Briefly, this model is based on
the density of states and it assumes that each nuclear-spin isomer
lies in its lowest rotational state. The branching ratios are computed
for exothermic channels (see Eq.~(13) of \cite{rist13}) and the
detailed balance principle is applied for the reverse endothermic
channels. The inclusion of reaction~(\ref{h3opeq}) in our network was
found to change the OPR of H$_2$O by less than 12\%. This small effect
reflects the fast destruction of H$_2$O by other abundant ions and
that of H$_3$O$^+$ by electrons, which both prevent H$_2$O from
efficiently exchanging protons with H$_3$O$^+$.

Finally, H$_3$O$^+$ could exchange protons with molecular hydrogen:
\begin{equation}
  {\rm pH_3O^+ + (o, p)H_2 \leftrightarrows oH_3O^+ + (o, p)H_2}.
  \label{h3opeq2}
\end{equation}
However, the reaction of H$_3$O$^+$ with D$_2$ was studied at 300~K
and isotope H/D exchange was not observed, with an upper limit for the
rate coefficient of $10^{-12}$~cm$^3$s$^{-1}$ \citep{kim75}. This was
interpreted as implying that the collision of H$_3$O$^+$ with H$_2$
does not form a stable intermediate complex, in contrast to the
reaction of H$_3$O$^+$ with H$_2$O. We can therefore {\it a priori}
neglect the thermalization of H$_3$O$^+$ by H$_2$, even if H$_2$ is
very abundant. A similar conclusion was reached by \cite{faure13}
regarding the similar NH$_4^+$ + H$_2$ reaction. In order to test the
potential impact of this reaction, however, we have computed
species-to-species rate coefficients by combining the above upper
limit ($10^{-12}$~cm$^3$s$^{-1}$) with the statistical model of
\cite{rist13}. Reaction~(\ref{h3opeq2}) was found to increase the OPR
of H$_3$O$^+$ by less than 1\% at 10~K and by about a factor of 2 at
50~K. It would be then desirable to theoretically investigate the
influence of hydrogen tunneling effects in the H$_3$O$^+$-H$_2$
complex.

In summary, the above equilibration reactions have a small or
negligible impact on the OPR of water and water ions. This is because
the destruction of these species is always faster than equilibration
processes. We emphasize, however, that this result holds only for
interstellar clouds where the atomic hydrogen and electron fractions
are small, i.e. lower than $\sim 10^{-1}$ and $\sim 10^{-6}$,
respectively.

%In diffuse environments, equilibration reactions with hydrogen atoms
%such as reaction~(\ref{h2opeq}) could play a significant role, as
%discussed by \cite{herbst15}.

%THE abundance of hydrogen atoms is such
%that if reaction~(\ref{h2opeq}) proceeds with a Langevin rate
%coefficient (i.e. $\sim 10^{-9}$~cm$^3$s$^{-1}$) then it will be
%faster than the (dominant) destruction reaction with electrons. If, in
%addition, the hydrogen density is so low that only the ground state of
%pH$_2$O$^+$ and oH$_2$O$^+$ is significantly populated, then the OPR
%of H$_2$O$^+$ will be thermalized at values between 1.4 and 1.8 in the
%temperature range 50-100~K. 

%\begin{figure}
%\includegraphics*[width=7.5cm,angle=-90.]{fig4.eps}
%\caption{OPRs of H$_3$O$^+$ and H$_2$O as function of kinetic
%  temperature. The solid lines give the thermalized OPRs. The thick
%  dashed lines correspond to our reference model. The thin dashed
%  lines correspond to our model including the thermalization reaction
%  H$_3$O$^+$+H$_2$O with the capture rate coefficients
%  $k_{cap}=2\times 10^{-9}$~cm$^3$s$^{-1}$ as well as $k_{cap}$
%  multiplied by a factor of 10, $10^2$ and $10^3$.}
%\label{fig5}
%\end{figure}

\subsubsection{Deuterated water}

It is interesting to investigate the OPR of deuterated water,
D$_2$O. As explained in Section~2.1, the chemistry of H$_2$O was
duplicated to include most of the deuterated homologue reactions. In
practice, the deuterium cloning was performed assuming that single
particle (H, H$^+$, D or D$^+$) hop is the dominant outcome of the
(complex forming) reactions, as in HB18 for the ammonia chemistry. For
the electronic dissociative recombination (DR) of D$_2$O$^+$ and of
the deuterated isotopologues of H$_3$O$^+$, we have adopted the rate
coefficients and branching ratios of H$_2$O$^+$ \citep{jensen99} and
H$_3$O$^+$ \citep{jensen00}, respectively. In addition, statistical
H/D branching ratios were assumed for the products. This latter
assumption is questionable because some DR experiments have suggested
the occurence of isotope (i.e. non-statistical) effects. Deviations
from statistical branching ratios are however generally small. An
exception is provided by the storage ring measurements of
\cite{jensen99} on HDO$^+$. These authors have shown that
recombination into OD + H is twice as probable as recombination into
OH+D, meaning that the release of H is favoured.

\begin{figure}
\includegraphics*[width=7.5cm,angle=-90.]{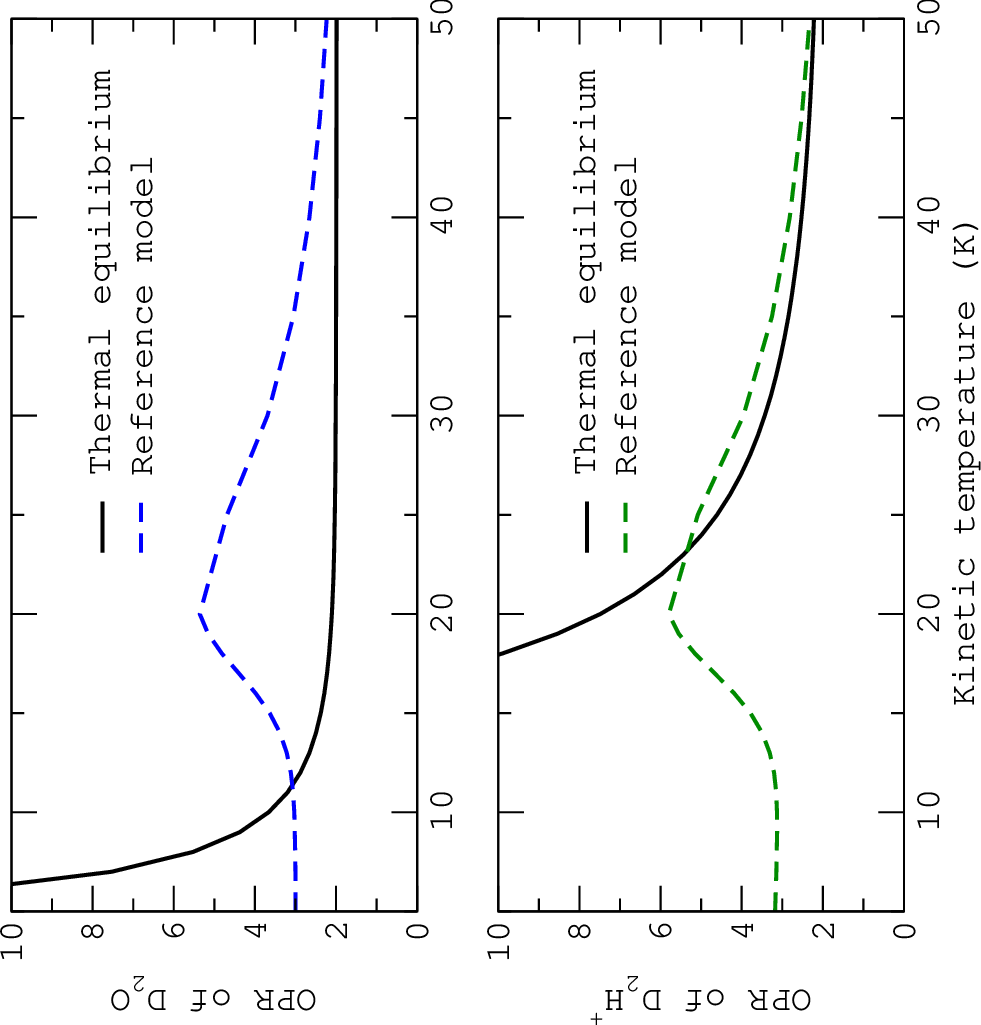}
\caption{OPRs of D$_2$O and D$_2$H$^+$ as function of kinetic
  temperature. The solid lines give the thermalized OPRs. The dashed
  lines correspond to our reference model. }
\label{fig4}
\end{figure}

The OPR of D$_2$O is plotted in Fig.~\ref{fig4} (upper panel) as a 
function of the kinetic temperature. It is shown to follow very
closely the OPR of D$_2$H$^+$ (lower panel), which is
quasi-thermalized down to 20~K due to the fast equilibration with
H$_2$. Thus, in contrast to H$_2$O whose OPR is controlled by the
spin-state of H$_2$, the OPR of D$_2$O is driven by that of
D$_2$H$^+$. This can be rationalized by considering that the gas-phase
formation of D$_2$O proceeds through the following reactions:
\begin{equation}
  {\rm O + D_2H^+ \to D_2O^+ + H},
  \label{d2op}
\end{equation}
\begin{equation}
  {\rm D_2O^+ + H_2 \to HD_2O^+ + H},
\end{equation}
\begin{equation}
  {\rm HD_2O^+ + e^- \to D_2O + H}.
\end{equation}
The corresponding nuclear-spin branching ratios are trivial since the
D$_2$ symmetry is conserved in these reactions. This implies that the
OPR of D$_2$O, HD$_2$O$^+$, D$_2$O$^+$ and D$_2$H$^+$ are strictly
equal. An important finding is thus that the OPR of D$_2$O should be a
good proxy for the OPR of D$_2$H$^+$.

Observationally, the OPR of D$_2$O was tentatively measured in the
cold envelope surrounding the protostar IRAS~16293-2422 by
\cite{vastel10}. The ground-state oD$_2$O and pD$_2$O lines were both
detected in absorption and an upper limit OPR(D$_2$O)$<$2.6 was
derived, which is consistent with our chemical model only if the gas
is warmer than 40~K. The analysis of these lines is however difficult,
in particular because the pD$_2$O transition has an emission
component. We note that the D$_2$O abundance was estimated as $\sim
10^{-11}$ \citep{vastel10}, which is within a factor of 2 of our
prediction for $T_{\rm kin}\sim 25$~K. Higher signal-to-noise ratio
observations are clearly needed to derive a more robust OPR. The
oD$_2$H$^+$ ground-state line was also detected in absorption towards
IRAS~16293-2422, but not the pD$_2$H$^+$ line \citep{harju17b}. Lower
limits OPR(D$_2$H$^+$)$>2.5$ and OPR(D$_2$H$^+$)$>1.7$ were derived by
\cite{harju17b} for the envelope and the ambient cloud,
respectively. These values are consistent with our model, but they do
not provide additional constraints on the kinetic temperature.  Future
observations should help to clarify the relation between these two
doubly deuterated molecules and to confirm - or disprove - the above
formation path.

\subsection{Larson-Penston simulation}

We now investigate the fractional abundances and OPRs of H$_2$O and
its ions during the gravitational collapse of a prestellar source of
initial mass $M_0=7$M$_{\odot}$. The Larson-Penston (L-P) simulation
assumes the same values of the cosmic-ray ionization rate
($\zeta=3\times 10^{-17}$~s$^{-1}$) and the initial radius of
refractory grain core ($a_g=0.1~\mu$m) as in the previous
calculations. The kinetic temperature is fixed at 10~K and the initial
density at 10$^4$~cm$^{-3}$. Other parameters can be found in Table~1
of HB18 (`reference' model). The steady-state abundances computed
above (as listed in Table~1 for some species) are used as the initial
composition of the collapsing sphere.

\begin{figure}
\includegraphics*[width=7.5cm,angle=-90.]{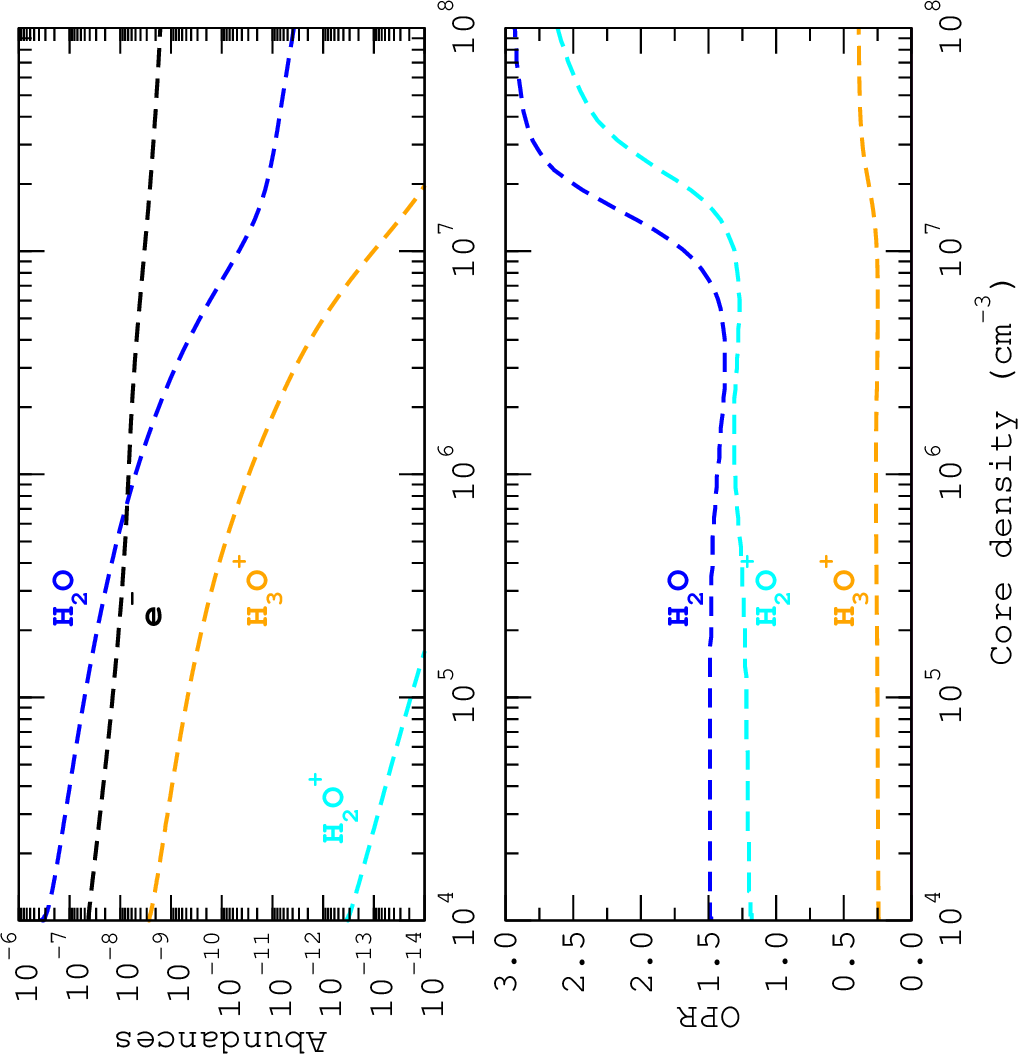}
\caption{Top panel: evolution of the fractional abundances of
  H$_3$O$^+$, H$_2$O, H$_2$O$^+$, and electrons - expressed relative
  to $n_{\rm H}$ - in a cloud that is undergoing contraction in a
  Larson-Penston (L-P) model. Bottom panel: evolution of the OPRs of
  H$_3$O$^+$, H$_2$O$^+$, and H$_2$O in the same L-P model.}
\label{fig5}
\end{figure}

At the onset of gravitational collapse, an envelope begins to form
around a core, as described in HB18. As the collapse proceeds, the
core contracts and its density increases. In Fig.~\ref{fig5} (upper
panel) are shown the variations in the fractional abundances of
H$_2$O$^+$, H$_3$O$^+$, H$_2$O and free electrons as functions of the
current density, $n_{\rm H}$, of the core (i.e. at the interface
between the core and the envelope since the core has a uniform
density). As the density increases, atoms and molecules are adsorbed
by the grains, whose radius (i.e. core plus ice mantle) increases as
more ice is deposited. We note that we assumed the same values $S=1$
of the sticking coefficient for all adsorbing species. In the upper
panel of Fig.\ref{fig5}, we observe a strong and similar decrease for
the abundances of H$_2$O, H$_2$O$^+$ and H$_3$O$^+$ as function of the
core density. This is due to the adsorption of the oxygenated species
by the grains. The slower reduction of the electron abundance reflects
the gas density increase ($n_e/n_{\rm H}\propto n_{\rm H}^{-0.5}$, see
\cite{flower07}). At a density of 10$^7$~cm$^{-3}$, the abundances of
H$_2$O and its ions have dropped by several orders of magnitude while
that of electrons is reduced by only a factor of 10. We also observe
that above a density of $\sim 10^7$~cm$^{-3}$, the decrease of the
H$_2$O abundance is appreciably flatter. This can be understood by
comparing the formation rates (per unit volume per unit time) of
H$_2$O due to the DR of H$_3$O$^+$ with that due to the cosmic-ray
(induced and direct) desorption of water ice . At a density $n_{\rm
  H}=10^4$~cm$^{-3}$, the former is $R^{DR}\sim 7.2\times
10^{-15}$~cm$^{-3}$s$^{-1}$ while the latter is $R^{des}\sim 3.5\times
10^{-17}$~cm$^{-3}$s$^{-1}$. At such low density, desorption is
negligible. At a density $n_{\rm H}=10^7$~cm$^{-3}$, however, we have
$R^{DR}\sim 1.9\times 10^{-14}$~cm$^{-3}$s$^{-1}$ and $R^{des}\sim
2.2\times 10^{-14}$~cm$^{-3}$s$^{-1}$ so that both formation pathways
are competing. At larger density, the formation rate of gas-phase
H$_2$O thus becomes dominated by the desorption of water ice. We have
also found that cosmic-ray induced photodesorption is the most
efficient process: it is typically a factor of 3 faster than
cosmic-ray direct desorption.

This change of regime is also observed in the lower panel of
Fig.\ref{fig5}: the OPR of H$_2$O is roughly constant and equal to 1.5
up to $n_{\rm H}$=10$^7$~cm$^{-3}$ where it smoothly increases to
reach the statistical value of 3, which is the OPR value assumed for
both UV photodesorption and cosmic-ray sputtering. A direct
consequence is that the main formation route of H$_2$O$^+$ at high
density is via the charge transfer H$_2$O+H$^+ \to$ H$_2$O$^+$+H. The
OPR of H$_2$O$^+$ is thus found to follow that of H$_2$O and to slowly
reach the value of 3. The OPR of H$_3$O$^+$ is, in turn, slightly
increased.

Observationally, the OPR of H$_2$O in prestellar cores is unknown
because only the ground-state oH$_2$O transition was detected in L1544
\citep{caselli12}. The total (ortho + para) H$_2$O column density was
computed by these authors assuming OPR(H$_2$O)=3. Since the central
density of L1544 is not larger than 10$^7$~cm$^{-3}$ \citep{keto14},
our model actually predicts that the OPR of H$_2$O should not exceed
1.5, implying that the column density derived by \cite{caselli12}
would be underestimated by about 20\%. We finally note that our model
predicts an H$_2$O abundance larger than 10$^{-9}$ for $n_{\rm
  H}\lesssim 3\times10^6$~cm$^{-3}$, which is in good agreement with
the estimate by \cite{caselli12}.

\section{Discussion}

Our spin-state chemical model predicts that the OPRs of H$_2$O$^+$,
H$_3$O$^+$ and H$_2$O in interstellar clouds lie in the range
$1.2-2.6$, $0.25-0.82$ and $1.5-2.6$, respectively. The ratios were
found to be constant below $\sim 30$~K and to depend essentially on
the spin-state of H$_2$. As noted above, this finding is valid for
molecular clouds where the H$_2$ fraction $f_{\rm H_2}$ is close to
unity and the electron fraction is lower than $\sim 10^{-6}$. We now
compare these results with available observational measurements.

%This result is based on the assumption that the gas-phase ion-neutral
%reactions~(2-6) all proceed via the formation of a long-lived
%intermediate complex in which a full scrambling of hydrogen atoms can
%take place.
%In prestellar cores where the central density exceeds this value,
%however, the OPR of H$_2$O was predicted to approach the statistical
%value of 3 because desorption of ices becomes the dominant source of
%gaseous H$_2$O in this regime.

Observational OPRs for H$_2$O$^+$ and H$_3$O$^+$ are scarce. To our
knowledge, there is only one {\it ISO} measurement for H$_3$O$^+$ in
the Sgr~B2 envelope where OPR(H$_3$O$^+)=0.8\pm 0.3$
\citep{goicoechea01}. This value is consistent with the statistical
value of unity, but it is also in agreement within error bars with our
prediction for $T_{\rm kin}\gtrsim 55$~K or OPR(H$_2)\gtrsim 0.4$ (see
Figs.~\ref{fig2}-\ref{fig3}). The OPR of H$_2$O$^+$ was determined in
more sources thanks to the {\it Herschel} satellite. In the diffuse
clouds towards the galactic center source Sgr~B2(M), it was found to
be almost constant at OPR(H$_2$O$^+)=3.2\pm 0.4$, which is consistent
with the statistical ratio of 3 \citep{schilke13}. \cite{herbst15} has
shown that this is also in agreement with the formation reaction~(4)
if this latter proceeds by hydrogen hopping rather than by
scrambling. He noticed that in such environment H$_2$O$^+$ could be
also the photoionization product of H$_2$O desorbed from ice
mantles. Similar values were derived in the diffuse clouds towards the
massive star-forming regions W49N and W31C with
OPR(H$_2$O$^+$)=$3.4\pm 0.6$ and OPR(H$_2$O$^+$)=$2.7\pm 0.4$,
respectively \citep{gerin13}. These values are again consistent with
the statistical ratio but the value in W31C is also compatible within
error bars with our prediction for $T_{\rm kin}\sim 80$~K or
OPR(H$_2)\sim 1$ (see Figs.~\ref{fig2}-\ref{fig3} and the discussion
below). For both water ions, higher signal-to-noise ratios would
clearly help to confirm or refute any deviation from the statistical
ratios.

\begin{figure*}
\includegraphics*[width=7.5cm,angle=-90.]{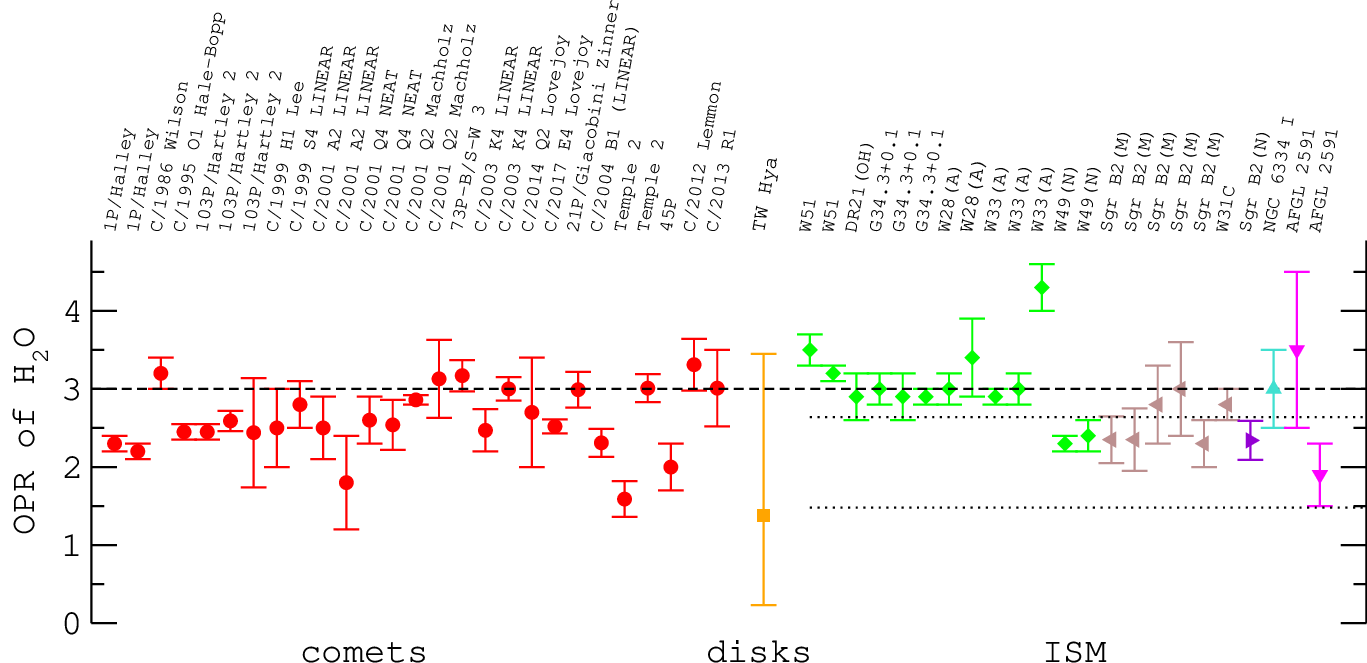}
\caption{OPR of H$_2$O as measured in comets, protoplanetary disks and
  the ISM. References are: Faggi et al. (2018) for comets; Salinas et
  al. (2016) for the protoplanetary disk TW~Hya (model Cm); Flagey et
  al. (2013) for W51, DR21(OH), G34.3+0.1, W28(A), W33(A) and W49(N);
  Lis et al. (2010) for Sgr~B2(M) and W31C; Lis et al. (2013) for
  Sgr~B2(N); Emprechtinger et al. (2013) for NGC~6334~I and Choi et
  al. (2015) for AFGL~2591. The black dashed-line gives the
  statistical value of 3. The two black dotted-lines correspond to our
  model at 10~K (OPR=1.48) and 100~K (OPR=2.64).}
\label{fig6}
\end{figure*}

Many more measurements are available for interstellar H$_2$O thanks to
{\it Herschel} observations. These are reported in Fig.~\ref{fig6}
(along with the OPRs derived for comets and the protoplanetary disk
TW~Hya). In the ISM, the OPR values for H$_2$O are generally
consistent with the statistical ratio of 3 within error bars. The
study of \cite{flagey13} is the most comprehensive. In total they
measured the water OPR for 13 translucent clouds. For these 13 clouds
the average OPR is $2.9\pm 0.1$. Of the 13 clouds, 10 have an OPR less
than 3$\sigma$ away from the statistical ratio. This is consistent
with the model of \cite{hollenbach09} where most of gaseous H$_2$O is
formed via photodesorption of water ice. One of the three other clouds
has an OPR value above 3 (towards W33(A)). The OPR of the remaining
two clouds is $2.3\pm 0.1$ and $2.4\pm 0.2$, as shown in
Fig.~\ref{fig6}. These two clouds are observed towards W49N, at
velocities $+40$~km~s$^{-1}$ and $+60$~km~s$^{-1}$, respectively. The
cloud at $+40$~km~s$^{-1}$ is also detected in NH$_3$ for which the
OPR is $0.5\pm 0.3$ , i.e. significantly lower than the statistical
value of unity \citep{persson12}. As noted above, \cite{faure13} have
shown that this low OPR is consistent with the nuclear-spin selection
rules for the formation of NH$_3$ in a para-rich H$_2$ gas. Similarly
here, the H$_2$O OPR of $2.3\pm 0.1$ is predicted by our spin-state
chemical model for OPR(H$_2)=0.6-1$ or $T_{\rm kin}=65-80$~K (see
Figs.~\ref{fig2}-\ref{fig3}). It is interesting to mention that
\cite{flagey13} have noted that this cloud is also detected in
H$^{13}$CO$^+$ at millimetre wavelengths, suggesting that its physical
properties approach those of dark clouds. A similar water OPR of
$2.35\pm 0.35$ was also measured in two clouds towards Sgr~B2(M), at
velocities $<-50$~km~s$^{-1}$ corresponding to the expanding molecular
ring \citep{lis10}. A third component probably blended with the Sgr~B2
envelope was also found with a low OPR of $2.3\pm 0.3$ but this was
attributed by \cite{lis10} to excitation effects. On the sight line
towards Sgr~B2(N), \cite{lis13} have also reported an average water
OPR of $2.34\pm 0.25$, in excellent agreement with the values found
for Sgr~B2(M). Finally, an even lower OPR of $1.9\pm 0.4$ was derived
by \cite{choi15} for the foreground clouds towards the high-mass
protostar AFGL~2591. Such a low value is in agreement with our model
for OPR(H$_2)<0.8$ or $T_{\rm kin}< 70$~K (see
Figs.~\ref{fig2}-\ref{fig3}). As noted by \cite{choi15}, taken
together these results show that water OPRs lower than 3 are found for
the translucent clouds with the highest column densities, i.e. in
regions where the interstellar FUV radiation field does not fully
penetrate and the physical properties are close to those of dark
clouds or dense cores.
%This is not
%yet confirmed by the OPR of the precursor ions H$_2$O$^+$ and
%H$_3$O$^+$ discussed above.

In summary, {\it Herschel} observations have shown that the OPR of
H$_2$O is statistical in diffuse/translucent clouds, which is
consistent with models where the formation rate of gas-phase H$_2$O is
dominated by the photodesorption of ice (Hollenbach et
al. 2009). These observations also indicate that the water OPR is
below the statistical value in some translucent/dense clouds and the
average ratio for these sources, OPR(H$_2$O)$\sim 2.3$, is consistent
with the ion-neutral nuclear-spin chemistry implemented in our model
(as denoted by the black dotted lines in Fig.~\ref{fig6}). It is also
possible that the FUV field is not entirely attenuated in these clouds
so that the formation rate of gas-phase H$_2$O is $\sim 50$\% via
gas-phase reactions (with an OPR of $1.5-2$) and $\sim 50$\% by
photodesorption of water ice (with an OPR of 3). The average water OPR
would then be $\sim 2.2-2.5$.

We can now compare these values with the water OPR reported for
planetary disks and solar system comets. To our knowledge, the first
OPR measurement in a propolanetary disk was reported by
\cite{pontoppidan10} for AS~205N using the Very Large Telescope. Their
best estimate was consistent with the high-temperature limit of 3 but
the low signal-to-noise ratio precluded a robust analysis. A better
measure was provided thanks to the {\it Herschel} observations of
TW~Hya by \cite{salinas16}. These authors have reported the value
OPR(H$_2$O)=$1.38^{+2.07}_{-1.15}$, which has large error bars but is
consistent both with the statistical ratio of 3 and the predictions of
our chemical model ($1.5-2.6$). More accurate observations are needed
to discriminate a statistical from a low OPR, which may provide an
important clue to the origin of water in disks.

Finally, a large number of measurements exist for comets, as compiled
recently by \cite{faggi18}. Fig.~\ref{fig6} shows the OPR of water
measured in 19 comets. The weighted-mean, $2.60\pm 0.03$, is
significantly lower than 3 and the corresponding spin temperature,
$\sim 31$~K, is significantly lower than the typical rotational and
kinetic temperatures in the coma \citep{bonev13}. The standard
deviation (0.03), however, was computed assuming uncorrelated
measurements and no systematic error. We note also that the median is
2.86. In any case, as discussed in the Introduction, the meaning of
the spin temperature is under debate. The current understanding is
that the OPRs in cometary comae are not indicative of the formation
temperature of ices, but instead reflect either the statistical value
(within uncertainties) or the gas-phase physical conditions in comae
(Hama et al. 2018). Observationally, no evidence for variation of the
OPR with depth in the nucleus or with nucleocentric distance in the
coma has been reported \citep{bonev13}. A possibility is that
gas-phase equilibration processes, which were found to be negligible
in dark cloud conditions, play a role in the very inner coma, as
discussed by \cite{shinnaka16}. In particular, the collisions of
H$_2$O with cold H$_3$O$^+$ ions could be an important
post-sublimation nuclear-spin conversion processes. In both scenarios
(statistical ratio or gas-phase processes), the spin-state of cometary
water would tell us nothing of the location and history of water
formation.

\section{Concluding remarks}

The nuclear-spin chemistry of oxygen hydrides was investigated using
the \texttt{UGAN} chemical network updated with the most recent
gas-phase kinetic data. The abundances and OPRs of gas-phase water and
water ions (H$_2$O$^+$ and H$_3$O$^+$) were computed under the
steady-state conditions representative of a translucent/dark molecular
cloud in a large temperature range (5-100~K). The predicted abundances
of OH$^+$, H$_2$O$^+$, H$_3$O$^+$ and H$_2$O were found in good
agreement with the `peak' abundances obtained by \cite{hollenbach09}
and \cite{hollenbach12} in their PDR model, i.e. at $A_V \sim 5$. The
OPRs of H$_2$O and its ions were found to deviate significantly from
both thermal and statistical values and to be entirely driven by the
OPR of H$_2$. The OPR of H$_2$O was shown to lie between 1.5 and 2.6
and to be consistent with values derived in translucent clouds with
extinction $A_V\gtrsim 3$. Calculations were extended to the early
phase of gravitational collapse of a prestellar core at 10~K using the
dynamical model presented in HB18. The direct and indirect cosmic-ray
desorption processes were found to control the abundance of gas-phase
water at densities $n_{\rm H}\gtrsim 10^7$~cm$^{-3}$, where the OPR of
H$_2$O increases from 1.5 to the statistical value of 3.

The main result of this work is that the low observational OPRs of
H$_2$O ($\lesssim 2.5$) measured in translucent clouds are consistent
with gas-phase ion-neutral chemistry {\it within the full scrambling
  hypothesis} and reflect a gas with OPR(H$_2$)$\lesssim 1$. Just like
the OPR of NH$_3$ \citep{faure13}, the OPR of H$_2$O (and also its
ions) therefore provides a diagnostic tool to study the `cold'
interstellar gas where H$_2$ is ortho-depleted (with respect to the
statistical value) and difficult to detect. This tool could prove
valuable in other environments such as extragalactic sources,
protoplanetary disks and comets. As discussed above, however, it
should be used as a probe of local physical conditions rather than
formation conditions.

More generally, the spectroscopy of ortho and para molecules in space
allows to study, perhaps uniquely, the nuclear-spin conservation of
identical nuclei in chemical reactions. Experimental evidence of
nuclear-spin selection rules is scarce and remains to be explored in
cold exothermic ion-neutral reactions, such as those involved in the
synthesis of H$_2$O. We note in this context the recent progress in
the control over the reactant quantum states in chemical
reactions. \cite{kilaj18} were able to (spatially) separate
ground-state oH$_2$O and pH$_2$O molecules which were reacted with
cold N$_2$H$^+$ in an ion trap (with a $\sim 20$\% higher reactivity
for pH$_2$O). The control over the quantum states of both reactants
and products is the next challenge. It will allow us to assess the
range of applicability of the scrambling assumption, on which our
results rely.

\section*{Acknowledgements}

This research was supported by the CNRS national program `Physique et
Chimie du Milieu Interstellaire'. Sara Faggi and Geronimo Villanueva
are acknowledged for providing the water OPRs for comets and for
useful comments. We also acknowledge Emmanuel Dartois, Marin Chabot
and Eric Quirico for helpful discussions about cosmic-ray sputtering
experiments. Finally we thank an anonymous referee for constructive
comments.

\appendix

\section{DIRECT COSMIC-RAY DESORPTION OF WATER}

The direct cosmic-ray desorption yield $\gamma_{\rm H_2O}$ (in
molecules~cm$^{-2}$s$^{-1}$) was computed by summing and integrating
the product of the sputtering yield $Y_s(\epsilon,Z)$ with the
differential cosmic-ray flux $j(\epsilon,Z)$:
\begin{equation}
  \gamma_{\rm H_2O} = 4\pi\sum_Z\int_{0}^{\infty}
  2Y_s(\epsilon,Z)j(\epsilon,Z)d\epsilon,
  \label{gamma}
  \end{equation}
where $\epsilon$ is the kinetic energy per nucleon, $Z$ is the atomic
number of the cosmic-ray nuclei and $Y_s(\epsilon,Z)$ is the
``sputtering'' yield in H$_2$O/ion obtained by combining the
experimental yield $Y_s(S_e)$ with the calculated electronic stopping
power $S_e(\epsilon,Z)$. It is multiplied by a factor of 2 to account
for the entrance and exit points of the cosmic-rays. From their
measurements and a compilation of data, \cite{dartois15} have fitted
the experimental yield $Y_s(S_e)$ as:
\begin{equation}
  Y_s(S_e)=\alpha S_e^{\beta},
\end{equation}
with $\alpha=4.4^{+4.3}_{-2.2}\times 10^{-3}$, $\beta=1.97\pm 0.07$
and $S_e^{\beta}$ is in units of ${\rm eV/10^{15}H_2O/cm^2}$. The
electronic stopping powers $S_e(\epsilon, Z)$ were computed with the
\texttt{SRIM-2013} code\footnote{http://www.srim.org.}
\citep{ziegler10} for a water ice density of 0.94~g~cm$^{-3}$ and for
the elements with the largest contributions, taking into account the
$\sim$ $Z^4$ scaling of the sputtering yield ($S_e$ varies as $\sim
Z^2$ at high-energy). This $Z^4$ dependence indeed largely compensates
for the low abundances of heavy ions (i.e. those with $Z\geq 6$). In
practice we thus included the contribution of 15 elements: H, He, C,
O, Ne, Mg, Si, S, Ca, Ti, V, Cr, Mn, Fe and Ni. Their fractional
abundances $f(Z)$ with respect to hydrogen (i.e. $f(1)=1$) were taken
from the Table~1 of \cite{kalvans16} (see references therein). We note
that iron is the species that gives the main contribution to the yield
(about 40-50\%).
%It is plotted in Fig.~\ref{iron} as function of the
%incident Fe ion energy for a water ice with density 0.94~g~cm$^{-3}$.

For the differential cosmic-ray flux $j(\epsilon,Z)$ (in
particles~cm$^{-2}$~s$^{-1}$~sr$^{-1}$~(MeV/amu)$^{-1}$) we adopted
the functional form proposed by \cite{webber83}:
\begin{equation}
  j(\epsilon,Z)=\frac{C(Z)E^{0.3}}{(E+E_0)^3},
  \label{flux}
\end{equation}
where $C(Z)=9.42\times 10^4 f(Z)$ is a normalizing constant and $E_0$
a form parameter which is between 0 and 940~MeV. The above formulation
corresponds to an initial or ``primary'' spectrum since it neglects
the energy-loss of cosmic-rays through the interstellar
gas. Nevertheless, it allows to explore the range of measured
ionization rates from diffuse to dense clouds (see Fig.~19 in
\cite{indriolo12}) by simply varying the parameter $E_0$. In addition,
we have found that the relation between $\gamma_{\rm H_2O}$ and
$\zeta$, as derived below, does not significantly depend on the
low-energy part of $j(\epsilon,Z)$.

\begin{figure}
\includegraphics*[width=7.5cm,angle=-90.]{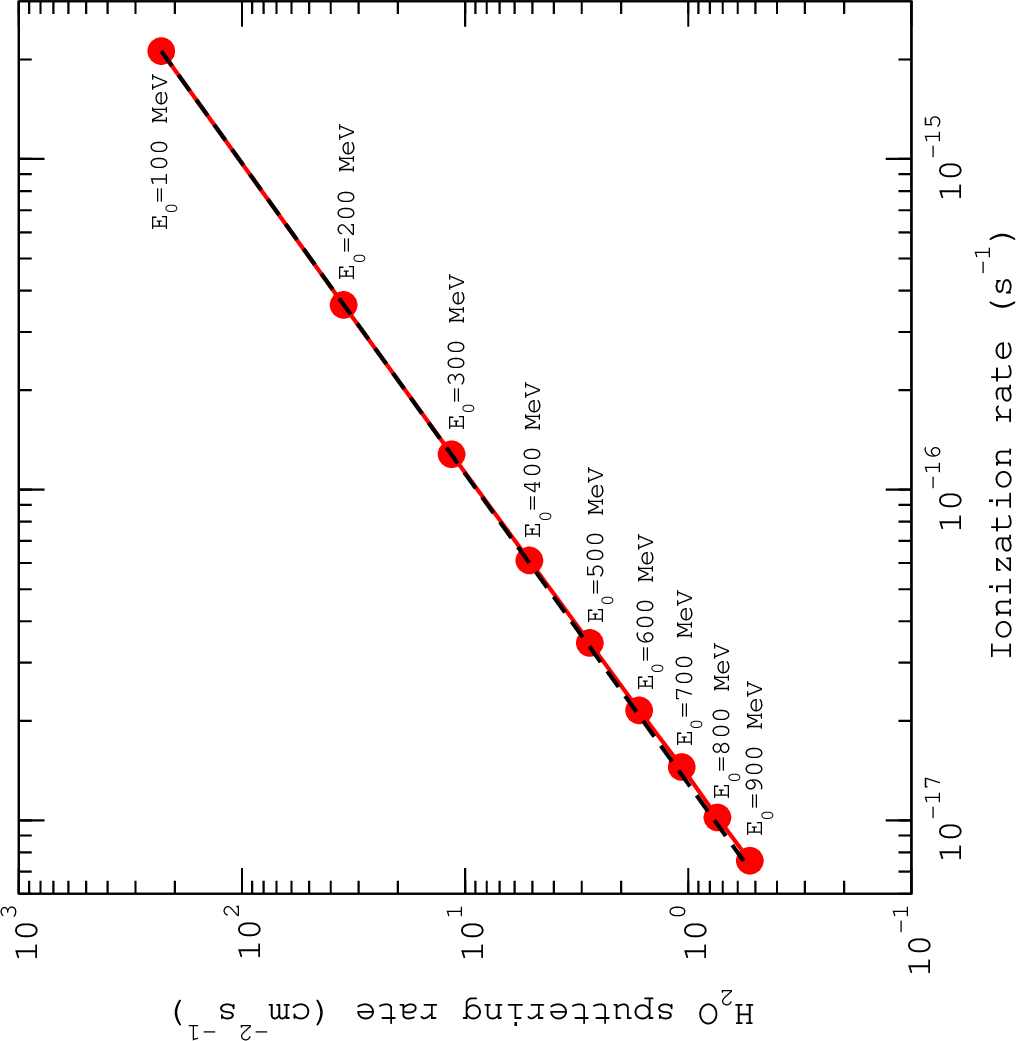}
\caption{Water sputtering rate $\gamma_{\rm H_2O}$ (in
  molecules~cm$^{-2}$~s$^{-1}$) as function of the ionization rate
  $\zeta$. The data points are computed from Eqs.~(\ref{gamma}) and
  (\ref{zeta}) with the form parameter $E_0$ varied from 100 to
  900~MeV. The dashed line gives a least-squares fit
  $a(\zeta/10^{-17}{\rm s^{-1}})^b$ with $a=0.76$ and $b=1.07$.}
\label{fig7}
\end{figure}

In order to infer the relation between $\gamma_{\rm H_2O}$ and
$\zeta$, it is necessary to compute the ionization rate $\zeta$:
\begin{equation}
  \zeta=4\pi\sum_Z\int_{0}^{\infty}(1+\Phi(\epsilon,Z))\sigma_{ion}(\epsilon,Z)j(\epsilon,Z)d\epsilon,
  \end{equation}
where $\sigma_{ion}(\epsilon,Z)$ is the ionization cross section and
$\phi(\epsilon,Z)$ is a correction factor accounting for the
contribution of secondary electrons to ionization. In the Bethe-Born
approximation, the ionization cross section only depends on the atomic
number $Z$ and the velocity of the incident particle so that
$\sigma_{ion}(\epsilon,Z)=Z^2\sigma_{ion}^{\rm p}(\epsilon)$ where
$\sigma_{ion}^{\rm p}(\epsilon)$ is the cross-section for ionization
of H$_2$ by proton impact. In addition, the secondary electron
contribution can be assumed identical for all elements and independent
of energy in the relevant range \cite[see][and references
  therein]{chabot16}. With the above approximations, the ionization
rate becomes:
\begin{equation}
  \zeta=4\pi(1+\eta)(1+\Phi)\int_{0}^{\infty}\sigma_{ion}^{\rm
    p}(\epsilon)j(\epsilon,1)d\epsilon,
  \label{zeta}
\end{equation}
where
\begin{equation}
  \eta=\sum_{Z\geq 2}f(Z)Z^2
  \end{equation}
is the correction factor for heavy nuclei ionization. The cross
section $\sigma_{ion}^{\rm p}(\epsilon)$ was taken from \cite{rudd85}
(see their Eqs.~(31-33) and Table ~III). For the secondary electron
contribution we adopted the correction factor $\Phi=0.7$
\citep{chabot16}. Finally $\eta=1.89$ was derived using the elemental
abundances of \cite{kalvans16}. It should be noted that all other
ionization processes, including electron capture, were neglected
because their contribution to the ionization rate was found to be less
than 1\% for the chosen cosmic-ray differential flux. In practice, the
$E_0$ parameter in Eq.~(\ref{flux}) was varied between 100 and 900~MeV
so that $\zeta$ was explored in the range $8\times 10^{-18} - 2\times
10^{-15}$. The integrals in Eqs~(\ref{gamma}) and (\ref{zeta}) were
evaluated numerically from $\epsilon_{\rm min}=100$~eV/amu to
$\epsilon_{\rm max}=10$~GeV/amu. It should be noted that the
low-energy part ($<100$~MeV) of the cosmic-cray spectrum is poorly
constrained \citep{chabot16} and leads to large uncertainties in the
ionization rate. On the other hand, we have found that the relation
between $\gamma_{\rm H_2O}$ and $\zeta$ is quite robust. The
calibration is plotted in Fig.~\ref{fig7} and one can notice that the
relation is very close to linear ($\gamma_{\rm H_2O} \propto
\zeta^{1.07}$). We suggest therefore to use the simple linear
relation:
\begin{equation}
  \gamma_{\rm H_2O}=0.8\left(\frac{\zeta}{10^{-17}{\rm
      s^{-1}}}\right),
  \end{equation}
for $\zeta$ in the range $\sim 10^{-17}-10^{-15}$~s$^{-1}$.

Finally, it is instructive to estimate the time scale for cosmic-ray
hits. Taking $E_0=600$~MeV (i.e. $\zeta=2.2\times 10^{-17}$s$^{-1}$)
with a fractional abundance of iron of $4.6\times 10^{-4}$, the time
between impacts for a grain of radius $a=0.1~\mu$m is:
\begin{equation}
  t_{\rm CR} = \left(4\pi^2
  a^2\int_{0}^{\infty}j(\epsilon,26)d\epsilon\right)^{-1}\sim 2\times
  10^4~{\rm yr},
  \end{equation}
which means that over the lifetime of a prestellar core ($\sim
10^6$~yr) a grain will experience about 50 impacts.

\section{ANALYTIC ORTHO-TO-PARA RATIO OF H$_2$O}

We derive below the OPR of water and water ions following their
formation via the reaction of OH$^+$ and H$_2$O$ ^+$ with H$_2$ and
the DR of H$_3$O$^+$ with electrons. We will assume {\it i)} that the
reactivities of {\it para-} and {\it ortho-} species are identical
(i.e. same overall formation and destruction rates) and {\it ii)} that
the destruction rates are faster than equilibration (thermalization)
processes. With these two hypotheses, the OPRs of H$_2$O$^+$,
H$_3$O$^+$ and H$_2$O are controlled by the formation paths. We will
also assume that all reactions proceed via the full scrambling of
protons in long-lived complexes.

Let us first consider the formation of oH$_2$O$^+$ and pH$_2$O$^+$
through the reaction of OH$^+$ with pH$_2$ and oH$_2$. Nuclear-spin
branching ratios can be derived using the approaches of \cite{oka04}
or \cite{quack77}:
\begin{equation}
  {\rm OH^+ + pH_2 \to} \left\{
\begin{array}{lr}
  \mbox{pH$_2$O$^+$ + H} &  \frac{1}{2} \\ \mbox{oH$_2$O$^+$ + H} &  \frac{1}{2}
\end{array}
\right.
\label{ohpph2}
\end{equation}
\begin{equation}
  {\rm OH^+ + oH_2 \to} \left\{
\begin{array}{lr}
  \mbox{pH$_2$O$^+$ + H} &  \frac{1}{6} \\
  \mbox{oH$_2$O$^+$ + H} & \frac{5}{6}
\end{array}
\right.
\label{ohpoh2}
\end{equation}
From the above equations we derive easily:
\begin{equation}
  {\rm OPR(H_2O^+)}=\frac{\frac{1}{2}+\frac{5}{6}{\rm
      OPR(H_2)}}{\frac{1}{2}+\frac{1}{6}{\rm OPR(H_2)}}.
\label{h2op}
\end{equation}
Likewise, by combining Eq.~(\ref{h2op}) with the nuclear-spin
branching ratios for the formation of H$_3$O$^+$:
\begin{equation}
  {\rm pH_2O^+ + pH_2 \to} \left\{
\begin{array}{lr}
  \mbox{pH$_3$O$^+$ + H} & 1 \\ \mbox{oH$_3$O$^+$ + H} & 0
\end{array}
\right.
\end{equation}
\begin{equation}
  {\rm pH_2O^+ + oH_2 \to} \left\{
\begin{array}{lr}
  \mbox{pH$_3$O$^+$ + H} &  \frac{2}{3} \\ \mbox{oH$_3$O$^+$ + H} &  \frac{1}{3}
\end{array}
\right.
\end{equation}
\begin{equation}
  {\rm oH_2O^+ + pH_2 \to} \left\{
\begin{array}{lr}
  \mbox{pH$_3$O$^+$ + H} &  \frac{2}{3} \\ \mbox{oH$_3$O$^+$ + H} & \frac{1}{3}
\end{array}
\right.
\end{equation}
\begin{equation}
  {\rm oH_2O^+ + oH_2 \to} \left\{
\begin{array}{lr}
  \mbox{pH$_3$O$^+$ + H} & \frac{1}{3} \\ \mbox{oH$_3$O$^+$ + H} & \frac{2}{3}
\end{array}
\right.
\end{equation}
we obtain:
%\begin{equation} {\rm
%  OPR(H_3O^+)=\frac{\frac{1}{3}OPR(H_2)+[\frac{1}{3}+\frac{2}{3}OPR(H_2)]OPR(H_2O^+)}{1+\frac{2}{3}OPR(H_2)+[\fr%ac{2}{3}+\frac{1}{3}OPR(H_2)]OPR(H_2O^+)}}, \end{equation}
\begin{equation}
  {\rm
    OPR(H_3O^+)=\frac{\frac{1}{6}+OPR(H_2)[\frac{7}{9}+\frac{11}{18}OPR(H_2)]}{\frac{5}{6}+OPR(H_2)[\frac{11}{9}+\frac{7}{18}OPR(H_2)]}}.
\label{h3op}
\end{equation}
By combining Eq.~(\ref{h3op}) with the branching ratios for the DR of
H$_3$O$^+$:
\begin{equation}
  {\rm pH_3O^+ + e^- \to} \left\{
\begin{array}{lr}
  \mbox{pH$_2$O + H} & \frac{1}{2} \\ \mbox{oH$_2$O + H} & \frac{1}{2}
\end{array}
\right.
\end{equation}
\begin{equation}
  {\rm oH_3O^+ + e^- \to} \left\{
\begin{array}{lr}
  \mbox{pH$_2$O + H} & 0 \\ \mbox{oH$_2$O + H} & 1
\end{array}
\right.
\end{equation}
we finally obtain:
\small
\begin{eqnarray}
{\rm OPR(H_2O)} & = & {\rm 2OPR(H_3O^+)+1} \\
& = & {\rm \frac{\frac{1}{3}+2OPR(H_2)[\frac{7}{9}+\frac{11}{18}OPR(H_2)]}{\frac{5}{6}+OPR(H_2)[\frac{11}{9}+\frac{7}{18}OPR(H_2)]}+1}
  \label{h2o}
\end{eqnarray}
\normalsize

The above Eqs~(\ref{h2op}), (\ref{h3op}) and (\ref{h2o}) are used in
Fig.~3 to plot the ``analytical'' OPRs as function of the OPR of
H$_2$. We note that if OPR(H$_2$)=3, these equations predict that
OPR(H$_2$O$^+$)=3, OPR(H$_3$O$^+$)=1 and OPR(H$_2$O)=3, as expected in
the statistical limit. On the other hand, if OPR(H$_2$) $\ll 1$, they
predict OPR(H$_2$O$^+$)=1, OPR(H$_3$O$^+$)=1/5 and OPR(H$_2$O)=7/5, as
observed in Fig.~3 (see the dotted lines).

As can be noticed in Fig.~3, the above equations slightly
underestimate the OPRs predicted by our reference model. This can be
explained by the reaction of O with H$_3^+$ which is another
(secondary) source of H$_2$O$^+$. It can be easily shown from the
following branching ratios:
\begin{equation}
  {\rm O + pH_3^+ \to} \left\{
\begin{array}{lr}
  \mbox{pH$_2$O$^+$ + H} &  \frac{1}{2} \\ \mbox{oH$_2$O$^+$ + H} &  \frac{1}{2}
\end{array}
\right.
\end{equation}
\begin{equation}
  {\rm O + oH_3^+ \to} \left\{
\begin{array}{lr}
  \mbox{pH$_2$O$^+$ + H} & 0 \\ \mbox{oH$_2$O$^+$ + H} & 1
\end{array}
\right.
\end{equation}
that the corresponding OPR of H$_2$O$^+$ is:
\begin{equation}
  {\rm OPR(H_2O^+)=2OPR(H_3^+)+1}
\end{equation}
Thus, at low OPR(H$_2$) where OPR(H$_3^+) \sim 0.4$ (see Fig.~1) this
reaction produces OPR(H$_2$O$^+$)$\sim 1.8$, which explains why the
actual OPR of H$_2$O$^+$ is slightly above unity. The OPRs of
H$_3$O$^+$ and H$_2$O are in turn slightly above 1/5 and 7/5,
respectively.

\bibliographystyle{mn2e}

\bibliography{faure-final}

\bsp

\label{lastpage}

\end{document}